\documentclass[preprint]{ptephy_v1}

\preprintnumber{none}

\usepackage{amsmath,amssymb}

\newcommand \ma \mathsf
\newcommand \tbt [4] { \begin{pmatrix}
    \ma #1 & \ma #2 \\ \ma #3 & \ma #4
  \end{pmatrix} }

\begin{document}

\title{Nambu-Goldstone modes in the random phase approximation}

\author{Kai Neerg\aa rd}
\affil{Fjordtoften 17, 4700 N\ae stved, Denmark}

\begin{abstract}%
  I show that the kernel of the random phase approximation (RPA)
  matrix based on a stable Hartree, Hartree-Fock, Hartree-Bogolyubov
  or Hartree-Fock-Bogolyubov mean field solution is decomposed into a
  subspace with a basis whose vectors are associated, in the
  equivalent formalism of a classical Hamiltonian homogeneous of
  second degree in canonical coordinates, with conjugate momenta of
  cyclic coordinates (Nambu-Goldstone modes) and a subspace with a
  basis whose vectors are associated with pairs of a coordinate and
  its conjugate momentum neither of which enters the Hamiltonian at
  all. In a subspace complementary to the one spanned by all these
  coordinates including the conjugate coordinates of the
  Nambu-Goldstone momenta, the RPA matrix behaves as in the case of a
  zerodimensional kernel. This result was derived very recently by
  Nakada as a corollary to a general analysis of RPA matrices based on
  both stable and unstable mean field solutions. The present proof
  does not rest on Nakada's general results.
\end{abstract}

\subjectindex{random phase approximation}

\maketitle

\section{Introduction}

The random phase approximation (RPA)~\cite{ref:Bo53} is ubiquituous in
many fields of physics including nuclear physics, and it is described
in textbooks such as the much cited monograph by Ring and
Schuck~\cite{ref:Ri80}. Formally it leads to the analysis of a matrix
$$
  \ma M = \tbt A B {{-B}^\ast} {{-A}^\ast} = \tbt 1 0 0 {-1} \ma S
$$
with
$$
  \ma S = \tbt A B {B^\ast} {A^\ast} ,
$$
where $\ma A$ and $\ma B$ are $n \times n$ matrices and $\ma A$ is
Hermitian and $\ma B$ symmetric so that $\ma S$ is Hermitian. The
matrix $\ma M$ is the \textit{RPA matrix} and $\ma S$ is the
\textit{stability matrix}. When $\ma M$ is constructed from
excitations of a Hartree, Hartree-Fock, Hartree-Bogolyubov or
Hartree-Fock-Bogolyubov self-consistent mean field solution, $\ma S$
is the Hessian matrix of the mean field energy with respect to
variations about self-consistency~\cite{ref:Th60-61}. This
mathematical problem is well analysed when $\ma S$ is \textit{positive
  definite}~\cite{ref:Th60-61,ref:Ri80}. Then $\ma M$ has $2 n$
linearly independent eigenvectors $\ma x_j$. The corresponding
eigenvalues $\omega_j$ form pairs of opposite nonvanishing reals and
the eigenvectors can be so normalised that %
$\ma x_j^\dagger \tbt 1 0 0 {-1} \ma x_k =%
(\text{sign\,} \omega_j) \delta_{jk}$.

It often occurs, however, that $\ma S$ is only positive
\emph{semidefinite}. Specifically, this is the case when the mean
field solution violates some continuous symmetry of the many-body
Hamiltonian, such as translational or rotational invariance, because
the mean field solution is then invariant to transformations within
the symmetry group. This leads to vanishing eigenvalues of $\ma M$,
and the number of linearly independent eigenvectors is generally less
than $2 n$. It is usually assumed that the eigenvectors corresponding
to such vanishing eigenvalues can be interpreted as associated, in the
language of classical analytic mechanics, with generalised momenta
whose conjugate coordinates describe local variations within the
symmetry group. These pairs of a cyclic coordinate and its conjugate
momentum form the so-called \textit{Nambu-Goldstone
  modes}~\cite{ref:Th60-61,ref:Nm60Go61}. However, it was not, to my
knowledge, until very recently proved that this interpretation is
always consistent with the structure of $\ma M$.

This situation changed due to work by Nakada, who presented an
extensive analysis of $\ma M$ in the most general case when no
definiteness of $\ma S$ at all is assumed~\cite{ref:Nk16}. In an
addendum to this work Nakada derives from his general formalism that
when $\ma S$ is positive semidefinite then the space acted on by %
$\ma
M$ is decomposed into three subspaces: one where the vectors of a
certain basis correspond to pairs of a coordinate and its conjugate
momentum that do not enter the Hamiltonian at all, one where they form
Nambu-Goldstone mode pairs and one where $\ma M$ acts as in the case
of a positive definite $\ma S$~\cite{ref:Nk16a}. I here give a proof
of this result which does not rest on Nakada's general formalism.

\section{Change of basis}

A unitary transformation gives
\begin{gather*}
  \ma S' = \tfrac12 \tbt 1 1 {{-i}} i \ma S \tbt 1 i 1 {{-i}}
    = \tbt C {E^T} E D , \\
  \tfrac12 \tbt 1 1 {{-i}} i  \tbt 1 0 0 {{-1}} \tbt 1 i 1 {{-i}}
    = \tbt 0 i {{-i}} 0
\end{gather*}
with \emph{real} matrices
$$
  \ma C = \Re \ma A + \Re \ma B , \quad
  \ma D = \Re \ma A - \Re \ma B , \quad
  \ma E = \Im \ma A + \Im \ma B .
$$
Because $\ma S'$ is a real, symmetric matrix, a further real,
orthogonal transformation maps it to a real, positive semidefinite,
diagonal matrix $\ma \Delta$. Applying both transformations
successively results in
\begin{equation}\label{eq:M''}
  \ma M'' = \ma N'' \ma \Delta ,
\end{equation}
where $\ma N''$ is imaginary and antisymmetric and obeys %
$\ma N''^2 = \ma 1$. Conversely any such matrix $\ma N''$ is mapped by
inverses of transformations of the above forms to $\tbt 1 0 0 {{-1}}$
and these transformations give, when applied to a real, positive
semidefinite, diagonal $\ma \Delta$, an $\ma S$ of the original
structure. As matrices of the form of $\ma M$ are thus unitary
equivalent to ones of the form of $\ma M''$, I drop the double primes
from now on.

\section{Onedimensional kernel of the stability matrix}

First assume for simplicity that $\ma \Delta$'s kernel is
onedimensional and its first diagonal element is zero while the rest
are positive. Let
$$
  \ma x^{(1)} = \begin{pmatrix} 1 \\ 0 \\ \vdots \\ 0 \end{pmatrix} .
$$
Due to the antisymmetry of $\ma N$ the real vector %
$-i \ma N \ma x^{(1)}$ belongs to the space orthogonal to %
$\ma x^{(1)}$, where orthogonality $\ma x \perp \ma y$ is defined by
$\ma x^T \ma y = 0$. It does not vanish because $\ma N^2 = \ma 1$.
With $\ma \Delta^{-1}$ denoting the diagonal matrix with first
diagonal element zero and the reciprocals of $\ma \Delta$'s diagonal
elements in the following positions (so it is not in a strict sense
$\ma \Delta$'s inverse, which does not exist), let
$$
  \ma x^{(2)} = a \ma \Delta^{-1} \ma N  \ma x^{(1)}
$$ 
with an imaginary normalisation factor $a$. Then
$$
  \ma M \ma x^{(1)} = 0 , \quad \ma M
  \ma x^{(2)} = a \ma x^{(1)} .
$$
When $a$ is chosen such that
\begin{equation}\label{eq:conj}
  \ma x^{(1)T} \ma N \ma x^{(2)} = -i ,
\end{equation}
we have $x^{(1)}_i = \{r_i,p\}$ and $x^{(2)}_i = \{r_i,q\}$, where $q$
and $p$ form a pair of a coordinate and its conjugate momentum and
$r_i, i = 1 \dots 2n$, are the canonical coordinates obeying
$\{r_i,r_j\} = -i N_{ij}$ in terms of which %
$\frac12 \sum_{i} \Delta_{ii} r_i^2$ is the Hamiltonian. Here
$\{.,.\}$ denotes the Poisson bracket; the quantal commutators
$[r_i,r_j]$ thus form the matrix $\ma N$. Because $\ma x^{(1)}$
belongs to the kernel of $\ma \Delta$, the Poisson bracket of the
Hamiltonian with $p$ vanishes, so $q$ is a cyclic coordinate. The
condition~\eqref{eq:conj} renders $a$ negative imaginary corresponding
to a positive mass. More precisely $i a$ is the reciprocal mass.

A canonical coordinate $r$ has vanishing Poisson brackets with these
two if and only if the vector $\ma x$ with coordinates %
$x_i = \{r_i,r\}$ satisfies
\begin{equation}\label{eq:res}
  \ma x^{(j)T} \ma N \ma x = 0, \quad j = 1, 2 .
\end{equation}
I call the space of such vectors the \emph{residual} space. Being the
orthogonal complement of $\text{span\,} (-i \ma N x^{(j)}, j = 1,2)$,
which is twodimentional because $-i \ma N$ is orthogonal and $\ma
x^{(1)}$ and $\ma x^{(2)}$ are linearly independent by mutual
orthogonality, the residual space has dimension $2n - 2$. The
relations~\eqref{eq:res} are easily seen to imply
$$
  \ma x^{(j)T} \ma N \ma M\ma x = \ma x^{(j)T} \ma\Delta \ma x = 0,
  \quad j = 1, 2 ,
$$
so the residual space is invariant to $i \ma M$. For $j = 1$ the
relation~\eqref{eq:res} requires that $\ma x$ is orthogonal to %
$-i \ma N \ma \ma x^{(1)}$. For $j = 2$ it is equivalent to
\begin{equation}\label{eq:x_1}
  \sum_k \ma N_{1 \cdot} \ma \Delta^{-1} \ma N_{\cdot k} x_k = 0 ,
\end{equation}
where $\ma N_{k \cdot}$ and $\ma N_{\cdot k}$ denote the $k$th row and
column vectors. Because the coefficient of $x_1$ in this equation is
positive the equation can be satified for any $x_k, k = 2, \dots, 2n,$
by adjustment of $x_1$. A basis for the residual space is thus
obtained by selecting a basis for the space orthogonal to %
$\ma x^{(1)}$ and $-i \ma N \ma x^{(1)}$ and supplementing each basic
vector by the first coordinate required by equation~\eqref{eq:x_1}.

Now let $\ma e^{(j)}, j = 1, \dots, 2n,$ be a real, orthonormal basis
for the total space such that
$$
  \ma e^{(1)} =\ma  x^{(1)} , \quad
  \ma e^{(2)} = -i \ma N \ma x^{(1)} ,
$$
and let, for $j = 3, \dots, 2n$, the vector $\ma x^{(j)}$ be obtained
by supplementing $\ma e^{(j)}$ by a first coordinate such as to
satisfy equation~\eqref{eq:x_1}. As these vectors are linearly
independent, they span the residual space. Moreover, because
$$
  \ma e^{(j)T} \ma x^{(k)} = \ma e^{(j)T} \ma e^{(k)}
  = \delta_{jk}, \quad  j,k \ge 3 ,
$$
the matrix
$$
  \sum_{j \ge 3} \ma x^{(j)} \ma e^{(j)T}
$$
performs the identity transformation within this space. In the residual
space the transformation $\ma M$ therefore has matrix elements
$$
  \ma e^{(j)T} \ma M \ma x^{(k)}
  = \sum_l \ma e^{(j)T} \ma N \ma e^{(l)}
           \ma e^{(l)T} \ma \Delta \ma x^{(k)} \\
  = \sum_{l \ge 3} \ma e^{(j)T} \ma N \ma e^{(l)}
                 \ma e^{(l)T} \ma \Delta \ma e^{(k)},
  \quad j,k \ge 3 ,
$$
where the reduction of the sum follows from
\begin{equation}\label{eq:x->e}\begin{gathered}
  \ma e^{(j)T} \ma N \ma e^{(1)}
    \propto \ma e^{(j)T} \ma e^{(2)} = 0 \quad
  \text{(or $\ma e^{(1)T} \ma \Delta \ma e^{(j)} = 0$)} , \\
  \ma e^{(j)T} \ma N \ma e^{(2)}
     \propto \ma e^{(j)T} \ma N^2 \ma e^{(1)}
     = \ma e^{(j)T} \ma e^{(1)} = 0, \quad j = 3 , \dots , 2n .  
\end{gathered}\end{equation}
These relations also give
\begin{multline*}
  \sum_{l \ge 3} \ma e^{(j)T} \ma N \ma e^{(l)}
    \ma e^{(l)T} \ma N \ma e^{(k)}
  = \sum_l \ma e^{(j)T} \ma N \ma e^{(l)}
    \ma e^{(l)T} \ma N \ma e^{(k)} \\
  = \ma e^{(j)T} \ma N^2 \ma e^{(k)}
  = \ma e^{(j)T} \ma e^{(k)} = \delta_{jk} , \quad j,k \ge 3 .
\end{multline*}
As the matrix $(\ma e^{(j)T} \ma \Delta \ma e^{(k)}, j,k \ge 3)$ is
positive definite and the matrix %
$(\ma e^{(j)T} \ma N \ma e^{(k)}, j,k \ge 3)$ is imaginary and
antisymmetric, the restriction of $\ma M$ to the residual space is
thus similar to a matrix of the form~\eqref{eq:M''} with a positive
definite $\ma \Delta$. Moreover, because
$$
  \ma x^{(j)T} \ma N \ma x^{(k)}
 = \left( \ma e^{(j)} + x^{(j)}_1\ma e^{(1)} \right)^T \\
     \ma N \left( \ma e^{(k)} + x^{(k)}_1\ma e^{(1)} \right)
 = \ma e^{(j)T} \ma N \ma e^{(k)} ,
  \quad j,k \ge 3 ,
$$
by the equations~\eqref{eq:x->e} and the antisymmetry of $\ma N$, the
matrix $-i (\ma e^{(j)T} \ma N \ma e^{(k)}, j,k \ge 3)$ is the matrix
of Poisson brackets of the canonical coordinates associated with the
basic vectors $\ma x^{(j)}, j = 3, \dots 2n$.

\section{Multidimensional kernel}

For a generalisation to the case when $\ma \Delta$'s kernel %
$\mathcal K$ has dimension $m$ greater than one, let %
$\ma e^{(j)},j = 1 \dots m,$ be orthonormal basic vectors for
$\mathcal K$ such that $\ma e^{(j)},j = m' + 1 , \dots m,$ span
$\mathcal K \cap i \ma N \mathcal K^\perp$. Assume %
$m' + 1 \le j \le m$. Then $\ma e^{(j)} \in i \ma N \mathcal K^\perp$
implies $\ma e^{(j+m-m')} \mathrel{\mathop:}=%
-i \ma N e^{(j)} \in \mathcal K^\perp$, and because $-i \ma N$ is
orthogonal, the latter vectors are orthonormal. The matrix $(\ma
e^{(j)T} \ma N \ma e^{(k)}, j,k \le m')$ is nonsingular. In fact, if
for some linear combination $\ma x$ of %
$ \ma e^{(j)}, j = 1 , \dots , m',$ the vector $-i \ma N \ma x$ would
be perpendicular to all $\ma e^{(k)}, k= 1 , \dots , m'$, then
because, by $\ma x \perp \ma -e^{(k+m-m')} = i \ma N \ma e^{(k)}$ and
the orthogonality of $-i \ma N$, the vector $-i \ma N \ma x$ is also
perpendicular to all $\ma e^{(k)}, k = m' + 1, \dots , m$, it would be
perpendicular to $\mathcal K$. But then $\ma x$ would belong to %
$i \ma N \mathcal K^\perp$, a contradiction. As %
$(\ma e^{(j)T} \ma N \ma e^{(k)}, j,k \le m')$ is also imaginary and
antisymmetric, it follows that $m'$ is even. The square of %
$(\ma e^{(j)T} \ma N \ma e^{(k)}, j,k \le m')$ is not necessarily the
unit matrix, but $(\ma e^{(j)T} \ma N \ma e^{(k)}, j,k \le m')$ can be
given this property by right and left multiplications by a nonsingular
square matrix and its transposed. This allows defining a basis for
$\text{span\,} (\ma e^{(j)}, j = 1 , \dots , m')$ whose vectors are
associated, in the manner detailed above, with pairs of a coordinate
and its conjugate momentum obeying the canonical Poisson bracket
relations (including vanishing of the Poisson brackets between
coordinates and momenta belonging to different pairs). These canonical
coordinates have vanishing Poisson brackets with the Hamiltonian, so
as parts of a complete set of pairs of a coordinate and its conjugate
momentum obeying the canonical Poisson bracket relations they will be
entirely absent from the Hamiltonian.

Let
\begin{gather*}
  \ma x^{(j)} = \ma e^{(j)} , \quad j = 1 , \dots , m , \\
  \ma x^{(j)} = \ma \Delta^{-1} \ma e^{(j)} , \quad 
 j = m + 1, \dots , 2m - m' ,
\end{gather*}
with $\ma \Delta^{-1}$ defined in the way analogous to that above.
These vectors are linearly independent because $\ma \Delta^{-1}$ is
nonsingular in $\mathcal K^\perp$. Like before the vectors $\ma
x^{(k)}, k = m' + 1, \dots , 2m - m'$ span a subspace invariant to %
$i \ma M$. Notice
$$
  \ma x^{(j)T} \ma N \ma x^{(k)} \propto \ma e^{(j)T} \ma e^{(k+m-m')}
  = 0, \quad m' + 1 \le j,k \le m .
$$
Because the matrix $\ma m$ of elements
\begin{equation}\label{eq:posdef}
  m_{jk} = i \ma x^{(j)T} \ma N \ma x^{(k+m-m')} \\
  = \ma e^{(j)T} \ma N \ma \Delta^{-1} \ma N \ma e^{(k)} ,
  \quad m' + 1 \le j,k \le m ,
\end{equation}
is positive definite, one can make a transformation among the vectors
$\ma x^{(j)}, j = m' + 1, \dots , m,$ to get
\begin{equation}\label{eq:canon}
  \ma x^{(j)T} \ma N \ma x^{(k+m-m')} = -i \delta_{jk} ,
  \quad m' + 1 \le j,k \le m .
\end{equation}
The transformation
$$
  \ma x^{(j)} \mapsto \ma x^{(j)} + \frac i 2 \sum_{k=m'+1}^m
    \ma x^{(k)} \ma x^{(k+m-m')T} \ma N \ma x^{(j)} , \\
  \quad j = m + 1 , \dots 2m - m'  ,
$$
then gives
$$
  \ma x^{(j)T} \ma N \ma x^{(k)} = 0,
  \quad m + 1 \le j,k \le 2m - m' ,
$$
without destroying the relations~\eqref{eq:canon}. After these
transformations one has
\begin{equation}\label{eq:dyn}
  \ma M \ma x^{(j)} = 0 , \quad
  \ma M \ma x^{(j+m-m')} = \sum_{k=m'+1}^m a_{jk} \ma x^{(k)} ,
  \quad j = m' + 1, \dots , m,
\end{equation}
where the matrix $i \ma a$ with elements $i a_{jk}$ is the inverse of
the matrix $\ma m$ defined by equation~\eqref{eq:posdef} \emph{before}
the transformations. As $i \ma a$ is symmetric and positive definite,
its appearance in equation~\eqref{eq:dyn} is rendered positive
diagonal by application \emph{after} the transformations of one more
orthogonal transformation simultaneously to both sets of vectors %
$\ma x^{(j)}$ and $\ma x^{(j+m-m')}, j = m' + 1 , \dots , m$. This
does not change %
$\ma x^{(j)T} \ma N \ma x^{(k)}, m' + 1 \le j,k \le 2m - m'$, and one
arrives at an interpretation of $\ma x^{(j+m-m')}$ and %
$\ma x^{(j)}, j = m' + 1 , \dots , m,$ as vectors corresponding to
pairs of a cyclic coordinate and its conjugate momentum. The negative
imaginary signs of %
$a_{jj}, j = m' + 1 , \dots , m,$ correspond to positive masses; in
fact the diagonal matrix elements of the transform of $\ma m$ by the
final orthogonal transformation are the masses themselves.

Now let the orthonormal set $(\ma e^{(j)}, j = 1 , \dots , 2m - m')$
be extended to an orthonormal basis for the entire space. Like before
the relations
$$
  \ma x^{(j)T} \ma N \ma x = 0, \quad j = 1,\dots , 2m - m',
$$
define another subspace invariant to $i \ma M$, the residual space.
These relations ensure that the canonical coordinates associated with
the two previous spaces have vanishing Poisson brackets with those
associated with the residual space. They are satisfied automatically
for $j = m' + 1 , \dots , m$ when $\ma x$ is a linear combination of %
$\ma e^{(k)}, k = 1 , \dots , m, 2m - m' +1 , \dots , 2n$.
Attempting to satisfy the remaining $m$ relations by supplementing one
$\ma e^{(k)}, k = 2m - m' +1 , \dots , 2n,$ by a linear combination of
$\ma e^{(l)}, l = 1 , \dots , m,$ gives $m$ linear equations, which
can be solved because %
$(\ma e^{(j)T} \ma N \ma e^{(k)}, 1 \le j,k \le m')$ and $\ma m$ are
nonsingular. This results in a basis %
$(\ma x^{(k)}, k = 2m - m' +1 , \dots , 2n)$ for the residual space.

Due to
$$
  \ma e^{(j)T} \ma x^{(k)} = \ma e^{(j)T} \ma e^{(k)}
  = \delta_{jk}, \quad  j,k \ge 2m - m' + 1 ,
$$
the matrix
$$
  \sum_{j \ge 2m-m'+1} \ma x^{(j)} \ma e^{(j)T}
$$
performs the identity transformation within the residual space, so the
restriction of $\ma M$ to this space has matrix elements
\begin{multline}\label{eq:product}
  \ma e^{(j)T} \ma M \ma x^{(k)}
  = \sum_l \ma e^{(j)T} \ma N \ma e^{(l)}
           \ma e^{(l)T} \ma \Delta \ma x^{(k)} \\
  = \sum_{l \ge 2m-m'+1} \ma e^{(j)T} \ma N \ma e^{(l)}
                 \ma e^{(l)T} \ma \Delta \ma e^{(k)},
  \quad j,k \ge 2m - m' + 1 ,
\end{multline}
where the reduction of the sum follows from
\begin{gather*}
  \ma e^{(l)T} \ma \Delta \ma e^{(k)} = 0 ,
  \quad l = 1 , \dots , m , \\
  \ma e^{(j)T} \ma N \ma e^{(l)}
     \propto \ma e^{(j)T} \ma N^2 \ma e^{(l-m+m')}
     = \ma e^{(j)T} \ma e^{(l-m+m')} = 0, \quad
       l = m + 1, \dots , 2m - m' .  
\end{gather*}
Because the vectors %
$\ma \Delta \ma x^{(j)}, j = 2m - m' + 1 , \dots , 2n,$ are linearly
independent and map to members of the residual space by the orthogonal
matrix $i \ma N$, the restriction of $\ma M$ to the residual space is
nonsingular. So is then the matrix %
$(\ma e^{(j)T} \ma N \ma e^{(k)}, j,k \ge 2m - m' + 1)$ by
equation~\eqref{eq:product}. This matrix is also imaginary and
antisymmetric. Its square is not necessarily the unit matrix, but %
$(\ma e^{(j)T} \ma N \ma e^{(k)}, j,k \ge 2m - m' + 1)$ can be given
this property by right and left multiplications by a nonsingular
matrix $\ma T$ and its transposed $\ma T^T$. Right and left
multiplications of the matrix %
$(\ma e^{(j)T} \ma \Delta \ma e^{(k)}, j,k \ge 2m - m' + 1)$ by %
$(\ma T^{-1})^T$ and $\ma T^{-1}$ then result in a similarity
transformation of the matrix %
$(\ma e^{(j)T} \ma M \ma x^{(k)}, j,k \ge 2m - m' + 1)$. As %
$(\ma e^{(j)T} \ma \Delta \ma e^{(k)}, j,k \ge 2m - m' + 1)$ is
symmetric and positive definite and this property is conserved by the
right and left multiplications by $(\ma T^{-1})^T$ and $\ma T^{-1}$,
the restriction of $\ma M$ to the residual space is thus similar to a
matrix of the form~\eqref{eq:M''} with a positive definite %
$\ma \Delta$.

Only when $m' = 0$ one has 
$$
  \ma e^{(j)T} \ma N \ma e^{(k)}
    \propto \ma e^{(j)T} \ma e^{(k+m)} = 0 , \\
  j = 2m + 1 \dots 2n , \ k = 1 , \dots , m ,
$$
so that
\begin{multline*}
  \sum_{l \ge 2m+1} \ma e^{(j)T} \ma N \ma e^{(l)}
    \ma e^{(l)T} \ma N \ma e^{(k)}
  = \sum_l \ma e^{(j)T} \ma N \ma e^{(l)}
    \ma e^{(l)T} \ma N \ma e^{(k)} \\
  = \ma e^{(j)T} \ma N^2 \ma e^{(k)}
  = \ma e^{(j)T} \ma e^{(k)} = \delta_{jk} , \quad j,k \ge 2m + 1 .
\end{multline*}
The transformations by $\ma T$ are then not required. Also
$$
  \ma x^{(j)T} \ma N \ma x^{(k)}
 = \ma e^{(j)T} \ma N \ma e^{(k)} ,
  \quad j,k \ge 2m + 1 ,
$$
because $\ma e^{(j)T} \ma N \ma e^{(k)}$ vanishes for %
$j = 1 , \dots , m , 2m + 1 , \dots , 2n$ and $k = 1 , \dots , m$. The
matrix $-i (\ma e^{(j)T} \ma N \ma e^{(k)}, j,k \ge 2m + 1)$ is then
the matrix of Poisson brackets of the canonical coordinates associated
with the basic vectors %
$\ma x^{(j)}, j = 2m + 1, \dots 2n$. For $m' > 0$ the matrix %
$-i \ma T^T (\ma e^{(j)T} \ma N \ma e^{(k)}, %
j,k \ge 2m - m' + 1) \ma T$ is not in general the matrix of Poisson
brackets of the canonical coordinates associated with the basic
vectors %
$\sum_{k \ge 2m - m' + 1} \ma x^{(k)} T_{kj}, j = 2m - m' + 1, \dots 2n$.

\section{Conclusion}

It was shown that when the random phase approximation (RPA) stability
matrix is positive semidefinite, the vector space on which it acts can
be decomposed into three parts: one where the vectors of a certain
basis correspond, in the equivalent formalism of a classical
Hamiltonian homogeneous of second degree in canonical coordinates, to
pairs of a coordinate and its conjugate momentum that do not enter the
Hamiltonian at all, one where they correspond to pairs of a cyclic
coordinate and its conjugate momentum (Nambu-Goldstone modes) and a
residual space where the RPA matrix acts as in the case of a positive
definite stability matrix. This was also proved very recently by
Nakada as a corollary to a general analysis of the most general RPA
matrix without limitations on the definiteness of the stability
matrix. The present proof does not rest on Nakada's general results.

\section*{Acknowledgment}

Discussions with Hitoshi Nakada are warmly appreciated.

\bibliographystyle{ptephy}\bibliography{goldstone}

\end{document}